\documentclass[5p,times]{elsarticle} 
\usepackage[english]{babel}

\usepackage{gensymb}
\usepackage{lipsum} 
\usepackage{algorithm}
\usepackage[normalem]{ulem}
\usepackage{graphicx} 
\usepackage{url}
\usepackage[flushleft]{threeparttable}
\usepackage{amsthm} 
\newtheorem{definition}{Definition}
\theoremstyle{definition}
\theoremstyle{remark}

\makeatletter
\def\eq{\@ifstar\@eq\@@eq}
\def\@eq#1{\begin{equation*}#1 \end{equation*}}
\def\@@eq#1#2{\begin{equation}\label{#1}#2 \end{equation}}
\makeatother

\usepackage{listings}
\usepackage{algpseudocode}
\usepackage{xargs}  

\usepackage{color,soul}

\usepackage{caption}
\usepackage{subcaption}
\usepackage[colorinlistoftodos,prependcaption, disable]{todonotes} 
\usepackage{amsmath}

\newcounter{todocounter}

\newcommandx{\blars}[1]{\textcolor{gray}{\lipsum[#1]}}

\usepackage{amssymb}

\biboptions{sort&compress,numbers}

\usepackage[figuresright]{rotating}

\begin{document}
\begin{frontmatter}

\title{Battery Life-Cycle Optimization and Runtime Control for Commercial Buildings Demand Side Management: A New York City Case Study}
\author[First]{Yubo Wang\corref{cor1}}
\ead{yubo.wang@siemens.com}
\author[Second]{Zhen Song\corref{cor2}}
\ead{zhen.song@siemens.com}
\author[Third]{Valerio De Angelis\corref{cor3}}
\ead{valerio@urbanelectricpower.com}
\author[First]{Sanjeev Srivastava\corref{cor1}}
\ead{sanjeev.srivastava@siemens.com}

\address[First]{Siemens Corporation, Corporate Technology. 755 College Road East, Princeton, NJ 08540.}
\address[Second]{Siemens Industry, Building Technologies Division. 9225 Bee Cave Road, Austin, TX 78733.}
\address[Third]{Urban Electric Power. 401 North Middletown Road, Pearl River, NY 10965.}

\fntext[label1]{This project is partially sponsored by NYSERDA under Agreement Number 39800.}

\begin{abstract}
In metropolitan areas populated with commercial buildings, electric power supply is stringent especially during business hours. Demand side management using battery is a promising solution to mitigate peak demands, however long payback time creates barriers for large scale adoption. In this paper, we have developed a design phase battery life-cycle cost assessment tool and a runtime controller for the building owners, taking into account the degradation of battery. In the design phase, perfect knowledge on building load profile is assumed to estimate ideal payback time. In runtime, stochastic programming and load predictions are applied to address the uncertainties in loads for producing optimal battery operation. For validation, we have performed numerical experiments using the real-life tariff model serves New York City, Zn/MnO$_2$ battery, and state-of-the-art building simulation tool. Experimental results shows a small gap between design phase assessment and runtime control. To further examine the proposed methods, we have applied the same tariff model and performed numerical experiments on nine weather zones and three types of commercial buildings. On contrary to the common practice of shallow discharging battery for preventing phenomenal degradation, experimental results show promising payback time achieved by optimally deep discharge a battery. 
\end{abstract}

\begin{keyword}
battery integration\sep commercial building\sep demand side management \sep stochastic programming.

\end{keyword}
\end{frontmatter}


\section{Introductions}

\subsection{Motivation}
\begin{figure}
\centering
\includegraphics[width=0.2\textwidth]{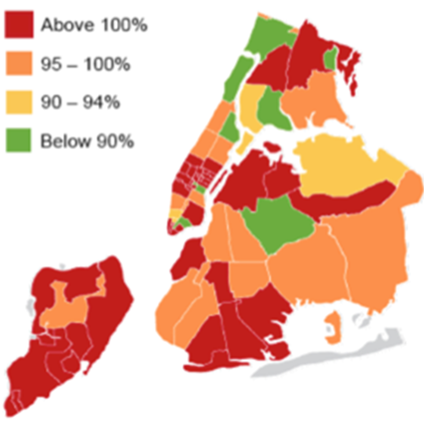}
\caption{Percentage utilization of the distribution network by 2018 without any demand response initiative as projected by ConEd (2009) ~\cite{CONED}}
\label{NYS_map}
\end{figure}

Metropolitan areas in U.S are facing stringent electric power supplies on peak demands. For example, Fig.~\ref{NYS_map} shows that in most of the locations of the New York City (NYC), the electricity peak demands are higher than the supply capacities, which is a direct result of continuously increasing loads, and closing of several coal and nuclear power plants due to economic and environmental concerns ~\cite{DOE_DR}. 

\begin{figure}[ht]
\centering
\includegraphics[width=0.35\textwidth]{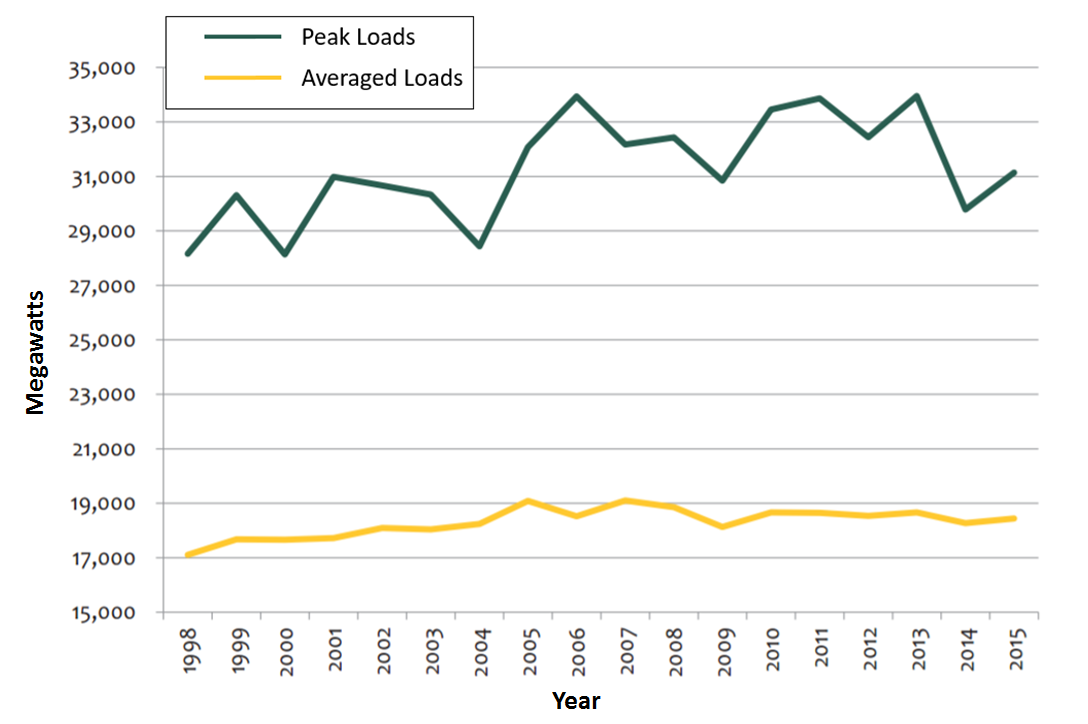}
\caption{Comparisons on New York State peak and averaged loads~\cite{NYISO_Trends2016}.}
\label{f1_peak}
\end{figure}

On the other hand, peak-to-average load ratio is an important indicator for efficiency operation of the grids. Fig.~\ref{f1_peak} demonstrates the comparisons of the historical New York State peak and averaged loads. It is shown that peak loads could be more than 80\% higher than averaged loads. It is projected the peak-to-average load ratio is even higher for NYC. A lot of generation capacities are built to meet the peak demands which last only for a short period of time.

Instead of enhancing generation side capacities, an alternative is to improve the Demand Side Management (DSM). It changes demands through various methods such as demand response, energy efficiency, customer-sited energy storage, and etc~\cite{ChiuDSM,alasseri2017review}. DSM  provides numerous benefits. First, there are environmental concerns to build more bulk generations, especially in the densely populated metropolitan areas. Second benefit lies in reduction of power systems investments. In 2014, ConEd has initiated Brooklyn-Queens demand management program to defer two substation upgrades that would otherwise cost \$1 billion. Instead of investments ConEd spent \$200 Million on behind the meter demand management and \$300 Million on other facility upgrades~\cite{RMI_ES17}. Third, DSM can reduce operation costs by avoiding frequently starting and stopping generators to meet peak demands which comes with a considerable cost~\cite{DOE_ES_DR16}. Although we are using NYC as an example, the value of DSM is applicable to other geographic landscapes as well. According to the DOE, up to 19\% of peak load, or 199 GW, of the USA may be shaved by 2019, if the maximum potential of DSM is fully implemented~\cite{FERC_03-12-10-DR}.

One market barrier in battery-based energy storage based DSM is end users are not likely to have positive payback for their battery investments~\cite{RMI_ES17, DOE_ES13}. Without a subsidy, the end users are unlikely to deploy batteries in large scale.  If end users can get payback in 3 to 5 years, the market penetration for customer-sited battery deployment may increase significantly, so that the pressure on the generation side will be relaxed significantly. 

\subsection{Literature Reviews}

According to ~\cite{LBNLReport}, DSM can be categorized into four classes, namely, Shape (days to years), Shift (hours to days), Shed (minutes to hours), and Shimmy (milliseconds to minutes). The scope of this study falls into the Shift and Shed timescales. Due to the delay in popular DSM protocols (OpenADR 2.0~\cite{openADR} and SEP 2.0~\cite{SEP20}), Shift and Shed are the applications tangible by the state-of-the-art technology, and is the focused scope of this paper. In regards to DSM strategy, both passive and active strategies exist. Passive DSM refers to DSM strategies that just utilizing the existing devices for DSM, for example through rescheduling the energy profile of HVAC. In contrast to passive DSM, active DSM utilizes new devices for load management, such as integration of energy storages. 

Rich literatures exist on passive DSM strategies customized for different building types: residential buildings~\cite{OzturkHomeDR}, general commercial building~\cite{ZhouDR, WangES16}, occupant-engaged commercial building DSM ~\cite{ZhenCDR16,yang2016building} and commercial building cluster DSM~\cite{SrivastavaDR15}, etc. There are several key differences for residential and commercial DSMs. First, DSM for residential buildings primarily studies control of home appliances based on the price signal. On the other hand,  DSM for commercial buildings uniquely involves controls of HVAC. Second, DSM in residential units use Time-of-Use (TOU) price as the incentive for end customers to shift their load. For commercial units, the loads are usually much larger and load peak becomes a concern. Critical Peak Price (CPP) kicks in on top of the TOU pricing.

Specifically in commercial buildings, the HVAC based DSM uses the building structure to provide short-term thermal storage capability. The DSM duration is about several hours, depending on the building. Researches showed that 10\% to 15\% load shaving is feasible without comfort loss or additional hardware installations~\cite{ZhenCDR16,yang2016building}. Despite the limitation, the advantage of this approach is that the operation cost is negligible.

Besides passively shifting the loads, many researchers have addressed active DSM through integrating battery based energy storages. For instance, batteries on Electric Vehicles (EVs) can be used for peak shaving and load balancing~\cite{ WangAE17, wang2017predictive}. Stationary battery pack can provide islanding capability, grid supports and economic operation~\cite{NottrottEnergyDispatch,mcpherson2018planning, 7286059}. The duration of battery based DSM is from hours to days, much longer than HVAC-based approach. However, the disadvantages of this approach are the installation and operational cost. 

In the center of battery based DSM is to leveraging the battery degradation cost. A handful of researches have been carried out in ~\cite{peterson2010,spanos2015life, ma2015distributed} for taking into consideration of battery degradation in DSM design. The model usually introduces non-convexity, thus making the DSM problems hard to solve.

Shifting our attentions to another critical component in DSM, the tariff model, literatures have primarily considered three pricing models: TOU ~\cite{yu2006optimal}, CPP ~\cite{liu2013data}, and the combination of the two ~\cite{wang2014adaptive}. The CPP models studied by the existing literatures are over-simplified problems that deviate far away from practical tariff models. Practical CPP model, for example the New York SC9 tariff model~\cite{SC9} used by this paper, usually calculates the peak demand based on time of the occurrence. The problem can be hardly modeled using standard convex optimization techniques, thus is hard to be solved. 

\subsection{Scope of This Paper}

In this paper, we consider an integrated DSM method, with the proposal of combining HVAC and battery control. To leverage the long load shifting capability of the battery and negligible operating costs of the HVAC system, we propose to charge the batteries at the time when the energy price and the demand are both below the average. Hours before the peak load, we could trigger HVAC system pre-cooling (pre-heating) mode for further load shaving. We consider a comprehensive price model, SC9 tariff model, used by ConEd. The price model is a combination of the TOU pricing and CPP pricing. And CPP pricing punishes the peak load by the time peak occurs. It is challenging to model this problem using convex optimization, thus we relax the problem and provide bounds for the problem.

Given there is no cost for HVAC based DSM, we first strive to answer two key questions in battery based DSM. 
\begin{itemize}
\item Design phase - What is the estimated payback time for a given battery taken into consideration the battery degradation cost?
\item Runtime - What is the control strategy to operate the battery in runtime, in order to achieve the payback time estimated by the previous question? 
\end{itemize}
The first problem is the key question to be answered before any battery is to be integrated. If the payback time is too long or if the battery will not secure positive payback time, it makes little sense to integrate battery. Note it is critical to consider battery degradation, otherwise, there is a considerable risk at damaging the battery way before the payback time. The second question arises from the fact that, the estimation of the payback time in the design phase is based on perfect knowledge of load profile. Therefore, the battery could precisely cut the loads of the billing cycle (especially the peak loads, which has a significant impact the utility bill). However, in reality, peak load in billing cycle is never perfectly known. Given that every discharge of the battery comes with a degradation cost, there is a good chance that if the operator wrongly operates the battery, it either misses the peak load of the billing cycle, or operate the battery too much that it significantly degrades the battery. To answer the two key questions, we propose two algorithms and validate its performance through numerical experiments. The design phase algorithm provides a best achievable payback time, and the runtime algorithm targets to ensure a payback time close enough to best achievable payback time

\subsection{Organizations}
The rest of the paper is organized as the followings. First, we present the system setup and the key technical challenge, the problem settings and formulate the problem into a bounded convex model. Then we present two algorithms. Algorithm 1 is for design phase battery life-cycle cost optimization, which estimate the payback time for a battery installation. Algorithm 2 is the runtime controller designed to cope with uncertainties in load profile predictions. We present the numerical validation results in Section 3 and conclude that it is feasible to achieve positive payback in 3 years.

\section{System Setup and Problem Formulation}\label{sec:arch}
\subsection{Challenge for Life-Cycle Cost Optimization}
    Life-Cycle cost optimization problem refers to determination of the payback time at design phase of battery integrations. One key challenge for such problem is the computation time. For comparison purpose, we have explored a heuristic-based optimization technique named Particle Swarm Optimization(PSO)~\cite{KennedyPSO07} , which has been used for numerous design phase problems~\cite{RashidiPSOSurvey}. The advantage of the algorithm is that it does not require analytical models and is applicable to very generic optimization, including non-convex problems. The disadvantage is its limited speed, especially for non-convex problem that this paper studies. At a 15 minutes simulation time step using EnergyPlus~\cite{EnergyPlus} and PSO, we estimate the minimum required time for solving this problem is over 21 years (using a Intel i5 16 G PC). This problem is then totally non-tractable.

\subsection{Algorithm Design and Data Flow}
\begin{figure}
\centering
\includegraphics[width=0.4\textwidth]{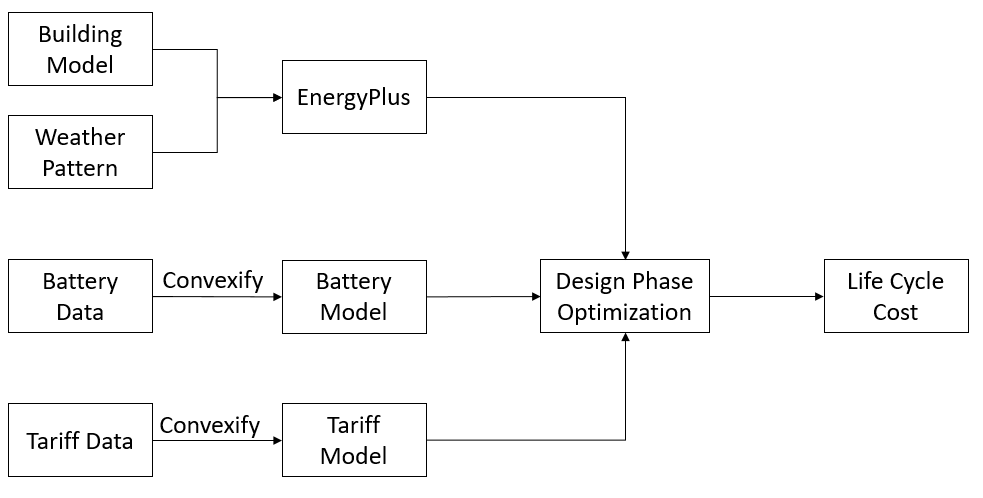}
\caption{Data flow diagram for life-cycle cost analysis (Algorithm 1).}\label{f:alg1}
\end{figure}

In order to quickly solve the life-cycle cost problem, we relax the original problem into a convex optimization problem~\cite{BoydBook}. In our case, we refer the life-cycle cost optimization algorithm as Algorithm 1, whose data flow diagram is shown in Fig.~\ref{f:alg1}. We first simulate the building energy profile in EnergyPlus using historical weather patterns and the extracted building profile. From experimental battery data, we developed a high fidelity battery simulation model and fitted it to a piecewise-linear analytical model. In order to minimize the discharge degradation of the battery, we use the battery only if the economic benefit of DSM exceeds its degradation cost. The economic benefit results from either shifting the energy profile (TOU pricing) or curtailing peak loads (CPP pricing), and is highly depend on the tariff model. In this paper we have relax the conditions of a highly non-convex SC9 tariff model into its convex counterpart.  With analytical and convex building, battery and tariff models, we can assess the whole battery life cycle in about 20 min, a much shorter time compared to using heuristic methods and non-convex model.

The aforementioned life-cycle cost in Algorithm 1 assumes the perfect knowledge on load profile in each billing cycle to curtail the peak loads. In reality, the performance is compromised by limited knowledge on load prediction for each billing cycle. The motivation of the Algorithm 2, i.e., runtime control, is to ensure that the payback of the real system is close to the Algorithm 1 as much as possible. Algorithm 2 and Algorithm 1 shares the battery modeling and tariff modeling parts. The major difference is on building loads. In Algorithm 2, based on forecasted weather pattern, the algorithm forecasts the monthly peak load and combine the results with historical loads. The detailed method is presented in our previous work~\cite{SrivastavaDR15}. Using stochastic programming framework, the runtime controller is able to achieve moving horizon control. The the runtime controller generates set points for the building automation system (BAS) and the battery management system (BMS).

\begin{figure}
\centering
\includegraphics[width=0.35\textwidth]{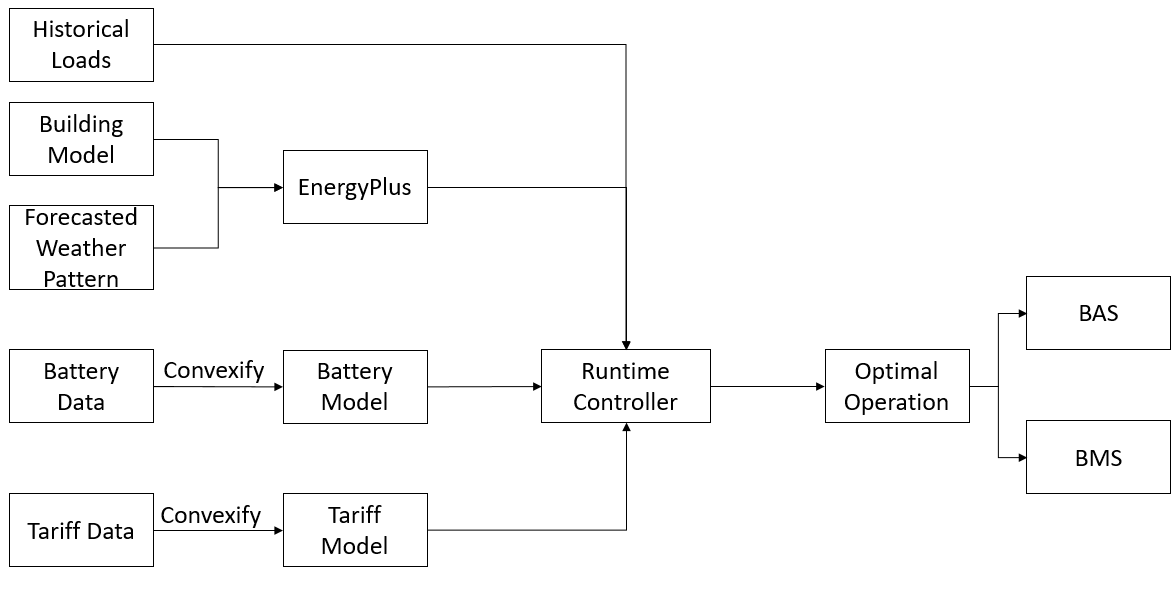}
\caption{Data flow diagram for runtime feedback control (Algorithm 2). }\label{f:alg2}
\end{figure}

\subsection{Tariff Model}\label{sec:tar}
\begin{figure}
\includegraphics[width=0.5\textwidth]{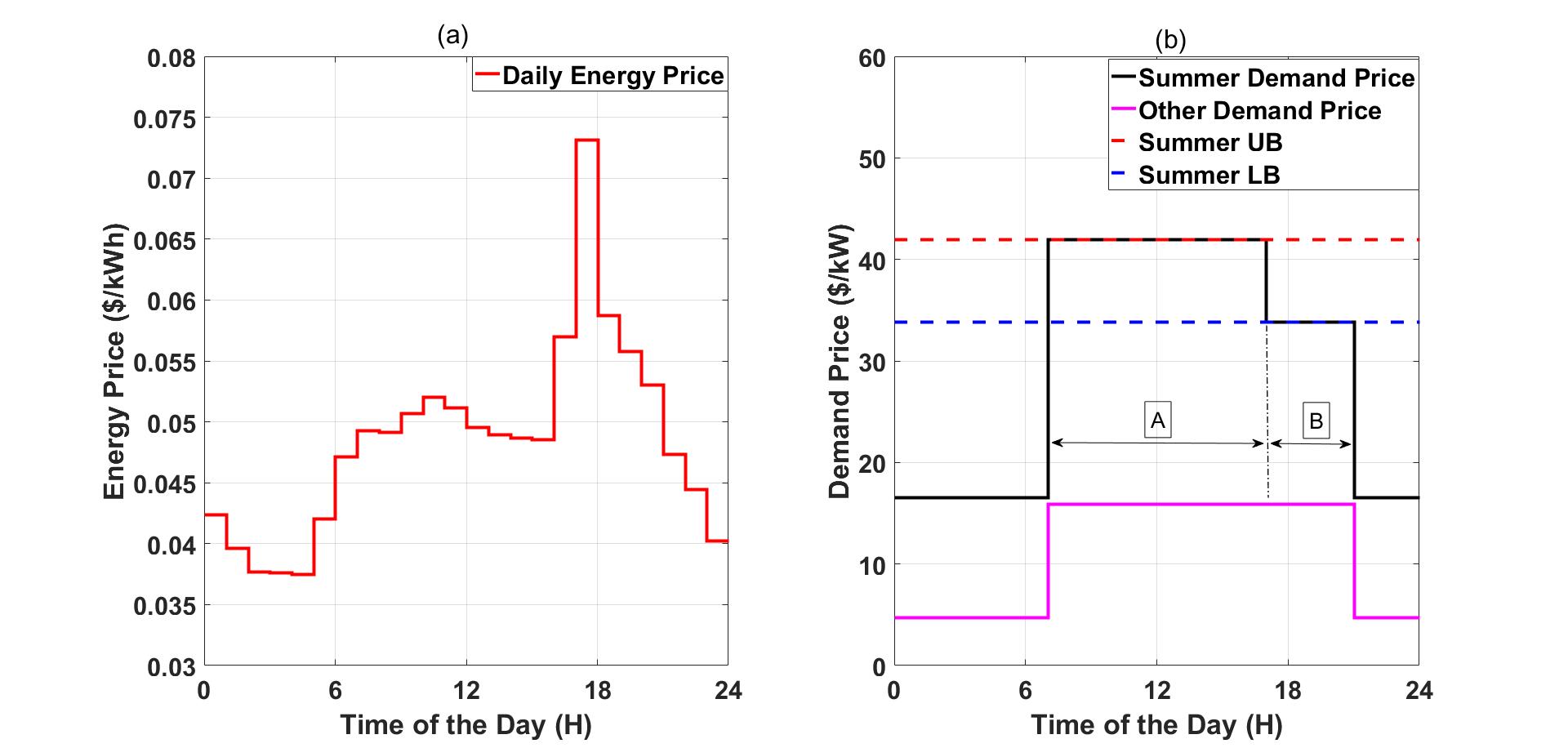}
\caption{Tariff for NYC. (a) energy price (b) demand price}
\label{fig:tariff}
\end{figure}

Tariff model stands for a set of rules how utilities charge their customers. It is the core of DSM, a strategy may change as tariff model varies. We use the real-life tariff model from ConEd as an example to study the DSM. The tariff model is plotted in Fig.~\ref{fig:tariff}. 

In this paper, we consider the SC9 tariff model with the billing cycle one calendar month. The length of the billing cycle is defined by the utility and is usually a calendar month. Though different tariff model has different details, they follow similar tariff patterns: the charge billed for a billing cycle is consist of \emph{energy charge} (TOU) and \emph{demand charge} (CPP). The energy charge is based on the amount of energy used in different time of a day throughout the billing cycle. On the other hand, demand charge is calculated based on the maximum power consumed by the customer throughout the billing cycle. Furthermore, according to the time that peak happens in the peak day, the demand charge can have the different price.We take the demand charge plot plotted in Fig.~\ref{fig:tariff}(b) to illustrate the demand charge concept. For example, it is in a summer month (June-September) and the peak throughout the month happens at 11 am, then it will be charged for \$41.95/kW for the peak power it consumed.

An important observation is that compared with unit energy charge, unit demand charge is considerably expensive. The demand charge can be a major part of the bill. Another inference could be made is the cost savings of DSM primarily comes from peak shaving (for demand charge) instead of load shifting (for energy charge). The numerical experiments carried out later have validated this inference.

\subsection{Battery Model}
\begin{figure}
\includegraphics[width=4.2cm,height=4cm]{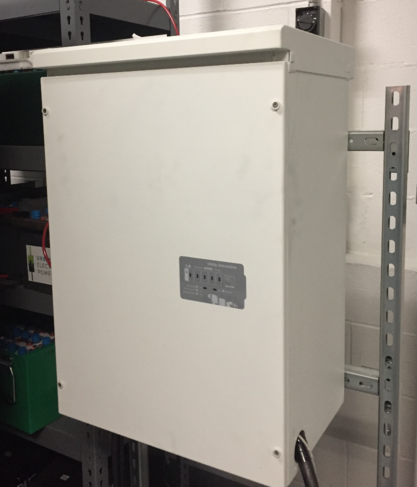} %
\includegraphics[width=4.2cm,height=4cm]{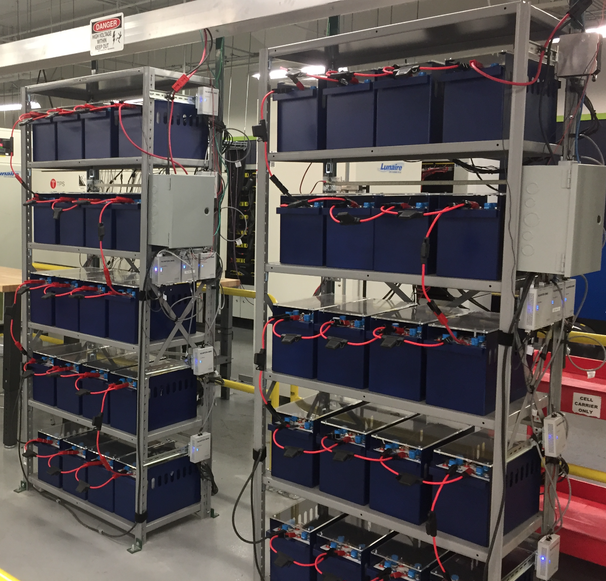}
\caption{UEP battery system. (Left: battery rack. Right: inverter)} \label{f:UEP}
\end{figure}

The battery performances are derived from the specifications of the Zn/MnO$_2$ prismatic cells from Urban Electric Power (UEP)~\cite{UEP}. The battery rack and it is inverter is shown in Fig.~\ref{f:UEP}. Users can monitor cell voltage distributions, state of health, and cells state of charge. The runtime controller is implemented as a component of Siemens Smart Energy Box~\cite{Lu2012Advanced}. The runtime controller can control BMS to change charging and discharging currents and query battery Depth of Discharge (DoD).

To better design DSM strategy that takes into account the degradation of battery, we study the battery characteristic in degradation. Note in \cite{zhou2011modeling}, it is shown that both DoD and ambient temperature rise could cause significant battery degradation. It further shows that the battery life-cycle decays quadratically with rise of temperature, while it decays exponentially with DoD. Thus, DoD is a more important consideration in modeling battery degradation. Together with many other factors not modeled, the battery degradation modeling in this paper only considers the major factor and therefore is not extensive. As DoD is a primary reason behind battery degradation, we need to extract the battery degradation for a discharge. 

\begin{figure}
\centering
\includegraphics[width=0.45\textwidth]{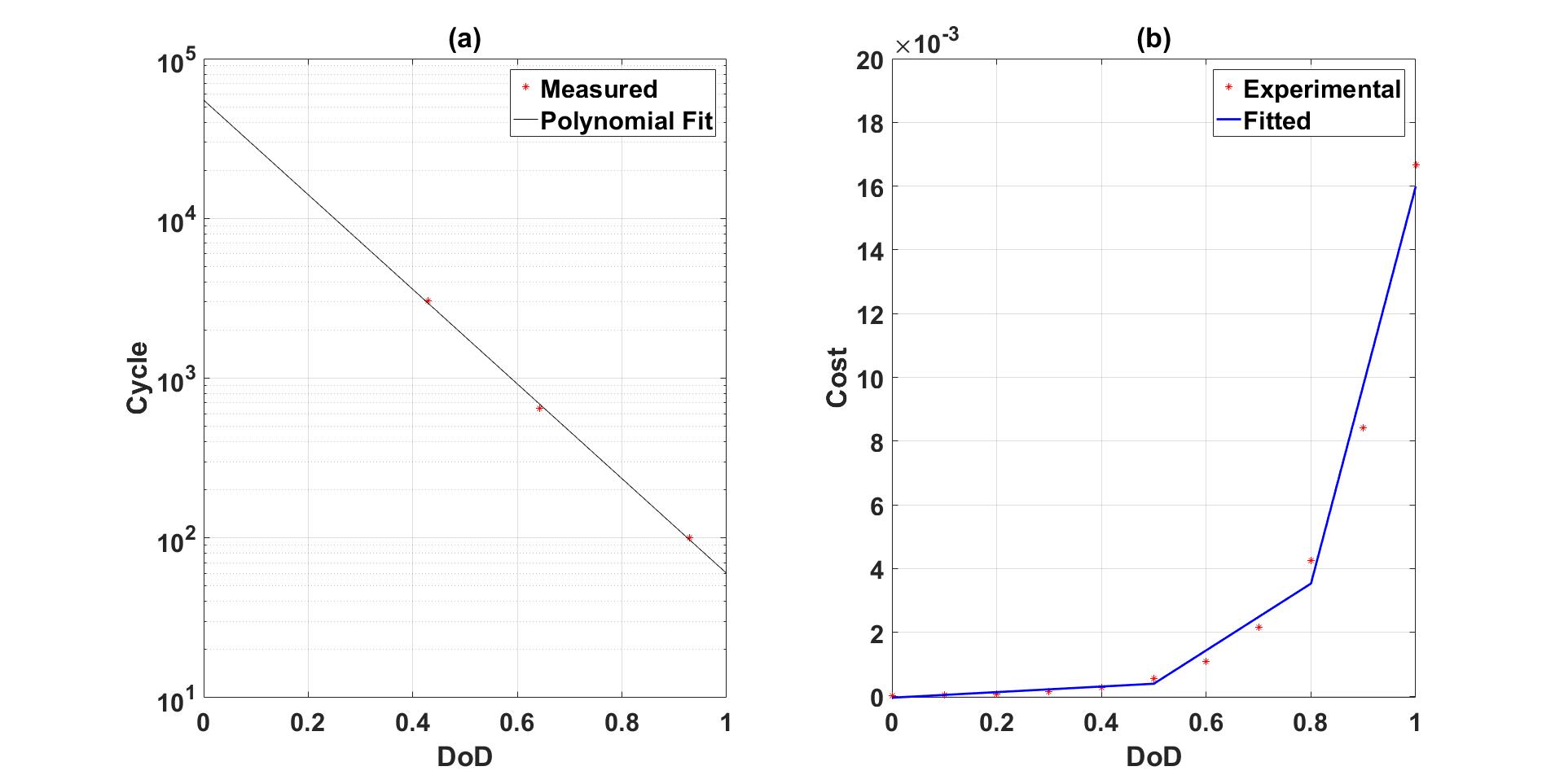}
\caption{Battery modeling (a) battery characteristic (b) battery degradation cost}
\label{fig:battery}
\end{figure}

Fig.~\ref{fig:battery}(a) plots experimental measurement of battery life cycle based on different DoD. The experiment is carried out on a new fully charged battery cell. It discharges the battery to certain DoD and recharges it back to full. After a number of cycles, the battery capacity degrades to the certain percentage of its initial capacity and therefore can no longer be used. The cycle number is defined as the life cycle of the battery under the DoD. Given the time it takes to plot one data point, three data points are collected. As depicted in the figure, it is observed that the battery degrades almost exponentially with DoD. 

We define the cost of a discharge from a fully charged battery , $C_d$, as follows:
\begin{equation}\label{eq:2.1}
C_d =C_b \frac{1}{\phi(D)} ,
\end{equation}
where $D$ is the DoD, $C_b$ is the battery cost and $\phi(\cdot)$ refers to the curve in Fig.\ref{fig:battery}(a) that maps DoD to its life cycle in log-scale. The logic behind this definition is intuitive: with $D$ corresponds to the life cycle $\phi(D)$, the cost should be an inverse of it times the initial cost of the battery. Following this definition, and assuming there is no degradation in charging, then a DoD $D_1$ to $D_2$ discharge will cost as follows:
\begin{equation} \label{eq:2.2}
 C_d=C_b[\frac{1}{\phi(D_2)}-\frac{1}{\phi(D_1)}]_+.
\end{equation}
Note that $[\cdot]_+$ stands for positive part of a function/scalar, i.e. $[x]_+=\max{\{0,x\}}$. Note that battery degradation could depend on its stage. The reasons behind using a degradation model independent of the battery stage is as follows: First, the time and economic cost of characterizing battery degradation at different stage is high. Second, precise modeling inevitably increases the complexity of the model, making it hard to run on resource-constraint controllers. Therefore, in this paper, we have considered a simplified stage independent degradation model also used in \cite{ortega2014optimal}.

We fit a line through the experimental data points in  Fig.~\ref{fig:battery}(a) using least square method. Based on the rather accurate fitted line, we create some more data points as depicted in Fig.~\ref{fig:battery}(b). The data points maps DoD to the cost of one discharge from fully charged status. The curve $\frac{1}{\phi(D)}$ that the data points depict is a concave function. However, the cost associated with discharging in Eq.~\ref{eq:2.2} is neither concave or convex. To accelerate the DSM control strategy, we convexify Fig.~\ref{fig:battery}(b) through a piece-wise linear function. The approximation is carried out using standard MILP as follows
\begin{equation}
\sum{w_j}=1,w_j \geq 0
\end{equation}
\begin{equation}
\begin{cases}
w_j \leq b_{j-1}+b_{j}, & \text{if $j \neq 1, end$}.\\
w_j \leq b_j, & \text{otherwise}.
\end{cases}
\end{equation}
\begin{equation}
DoD_{x}w= SoE_{ini}+\sum_t{P_{bat}^i}
\end{equation}
\begin{equation}
C_{deg}^i = Cost_{y}w
\end{equation}

where $w_j \in \mathbb{R}^+$ and $b_j \in \{0,1\}$ are ancillary variables. Assuming there are $H$ time slots in a day, $P_{bat}^i \in \mathbb {R}^H$ represents the battery power in day $i$. $DoD_{x} \in  \mathbb {R}^D$ and $Cost_{y} \in  \mathbb {R}^D$ correspond to the x-axis and y-axis value of $D$ data points in Fig.\ref{fig:battery}(b). $SoE_{ini}$ and $SoE_{max}$ stand for the initial and maximum State-of-Energy (SoE) of the battery. $b_j$ is an ancillary decision variable to $w_j$ and acts as activations function for $w_j$. Note as the sum of $w_j$ is 1. Therefore, only one $b_j$ can be one, others have to be zeros. The combination of $w_j$ and $b_j$ allows using piece-wise linear approximations for the curve in Fig.\ref{fig:battery}(b).

\subsection{Algorithm 1: Design Phase Battery Life-Cycle Optimization}
Before any battery installation, it is important to optimize the battery and inverter size, in order to minimize the life-cycle costs. Since we assume perfect knowledge on the demand load, the calculated life-cycle is shorter than reality. The life-cycle optimization, i.e., Algorithm 1, suggests a theoretical limit for the shortest payback time. The real world runtime controller, although can have different designs, will not provide shorter payback time than what Algorithm 1 predicts.

Following the tariff model described in Section 2.3, assuming there are $H$ time slots in one day, the control objective of the design phase assessment tool will be:
\begin{equation} 
\begin{aligned}
& {\text{min}}
& & \sum{(C_{deg}^i+P_E(P_{bat}^i+P_{load}^i))}+P_D\underset{i,t}{\text{max}}\{P_{bat}^i+P_{load}^i\} \\
\end{aligned}
\end{equation}
where $C_{deg}^i \in \mathbb R $ is the battery degradation in day $i$ in a billing cycle. $P_E \in \mathbb {R}^H$ stands for the price of the day for energy usage. $P_{load}^i \in \mathbb {R}^H$ is the fixed load profile throughout day $i$. $P_D \in \mathbb {R}$ is the demand price. 

The battery constraints are modeled as follows:
\begin{equation}
P_{min}\leq P_{bat}^i\leq P_{max}, \forall i
\end{equation}
\begin{equation}
0\leq \sum_t{P_{bat}^i}+SoE_{ini} \leq SoE_{max}, \forall i
\end{equation}
\begin{equation}
\sum_{t=H}{P_{bat}^i}+SoE_{ini} = SoE_{ini}, \forall i
\end{equation}
where $\sum_t$ is the summation of the first $t$ component. The constraints on the batteries limit the power of the charging/discharging power, and prevent controllers from over-charging or over-discharging of the battery. Furthermore, after a full day's operation, the battery should be reset to its initial status for it to get ready for the next day's operation.

As is shown in the Fig.\ref{fig:tariff}(b), the demand price is based on time of use. It is hard to model the price in a convex setting, which limits the efficiency in problem solving. Without detailed prove, we use the upper and lower bound price shown in Fig.~\ref{fig:tariff}(b) to as a relaxed demand price. We will solve the assessment problem twice, getting a range of payback time. However, every time the problem is solved, it needs to be verified whether the demand load locates at 7 am to 8 pm (peak time).Given the load profile, we have not yet met the circumstance that the demand load locates at off-peak time.

Then the design phase assessment tool is formulated as:
\begin{alignat*}{2}
\min_{P_{bat},w,b} \quad & \sum{(C_{deg}^i+P_E(P_{bat}^i+P_{load}^i))}+P_D\underset{i,t}{\text{max}}\{P_{bat}^i+P_{load}^i\} \\
  \text{s.t.} \quad &  \begin{aligned}[t]
	\text{(4)-(10)}
  \end{aligned}
\end{alignat*}

\subsection{Building Energy Model}
\begin{figure}
\centering
\includegraphics[width=0.35\textwidth]{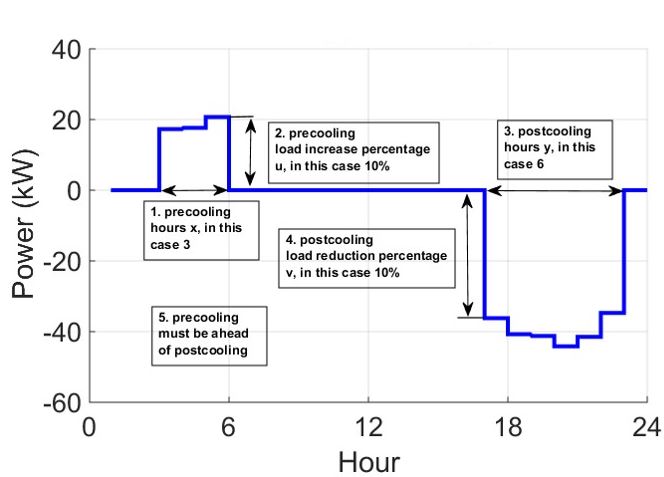}
\caption{HVAC pre-cooling and post-cooling}
\label{fig:building}
\end{figure}

According to our previous research in~\cite{ZhenCDR16}, building could be pre-cooled before peak load. The concept is illustrated in Fig.\ref{fig:building}. There are additional constraints added to the original problem. Firstly, before the peak load, there are are few hours, characterized by $x$ in the figure, for pre-cooling. The load increase percentage is $u$ percent. Second, in the post cooling period, the HVAC system could be partially shutdown. Accordingly it results in a load decrease of $v$ percent for $y$ hours. Finally, the additional constraint is pre-cooling must happen before post-cooling. 
When the HVAC control is taken into account, the new objective function will be the following:
\begin{equation}
\begin{aligned}
 {\text{min}} &\sum{(C_{deg}^i+P_E(P_{bat}^i+P_{load}^i+P_{pre}^i+P_{post}^i))} \\
&+P_D\underset{i,t}{\text{max}}\{P_{bat}^i+P_{load}^i+P_{pre}^i+P_{post}^i\}
\end{aligned}
\end{equation}
$P_{pre}^i \in \mathbb {R}^H$ and $P_{post}^i \in \mathbb {R}^H$ stand for the pre-cooling power and the post-cooling power. Additional constraints on HVAC can be implemented representing the logics shown in Fig.\ref{fig:building}.

\subsection{Algorithm 2: Runtime Controller}
The runtime control loop does not assume perfect knowledge on load profile. Consequently, using the same tariff model cannot capture the peak load in the billing cycle and cut it accordingly. To resolve this problem, we use a moving horizon stochastic controller to account for the uncertainties in one billing cycle. The objective function for runtime can be described as follows:
\begin{equation}
\begin{aligned}
{\text{min }} &C_{deg}^i+C_{deg}^F(\eta)+P_E(P_{bat}^i+P_{load}^i) \\
&+P_D\underset{i,t}{\text{max}}\{P_L^H,P_{bat}^i+P_{load}^i,P_L^F(\eta)+P_{bat}^F(\eta)\} 
\end{aligned}
\end{equation}
in which $\eta$ stands for the uncertainty to the end of the billing cycle. $C^F_{deg} \in \mathbb {R}$ is the future battery degradation. $P^H_L \in \mathbb {R}^H$ corresponds to the historical peak load day $P^F_L \in \mathbb {R}^H$ is the future peak load day. $P_{bat}^F \in \mathbb {R}^H$ stands for the battery operation in the future peak day. The maximum function leverages the recorded historical load, the peak load of the operational day. And the future peak loads in the current billing cycle. Note we do not account for the uncertainties in the operational day. The reason on one end, is it dramatically increase the complexity of the algorithm, and on the other end, the load in the operating day can be better predicted with minor errors. Instead of optimizing through one full billing cycle in Algorithm 1, we optimize over one day in runtime controller. The optimization horizon moves as the time passes. 

The question then is how to model the uncertainty of the future peak load.  We use Kernel Density Estimation (KDE) to capture the stochastic variables. KDE is a model-free density estimation. Then KDE is used to generate samples of the stochastic variables. Based on the generated samples (called scenarios), Monte Carlo simulations are run to approximate the real distribution of stochastic variable. This process is named sample averaged approximation. For a detailed discussion, please refer to ~\cite{WangES16}.

\section{Results and Analysis}
This section validates the proposed DSM strategy and compares the proposed DSM strategy with state-of-art technique. To further study the effectiveness of the proposed DSM, case studies are carried out for different geographical areas and commercial building types.
\subsection{Numerical Environmental Setup}
In this section, we validate the proposed DSM using real-life data. The tariff of ConEd is used for evaluation and the model is plotted in Fig.~\ref{fig:tariff}. For weather data input to EnergyPlus, NYC data of the year 2015 is used. We used the large office in the DOE reference commercial building category as the benchmark test of this numerical study. The proposed DSM problem is solved with MOSEK on a PC with 3.4GHz CPU and 16G memory.

\subsection{Numerical Case Studies}
In this section, we run numerical studies over three different DSM strategies in, namely DSM neglecting battery degradation, DSM considering battery degradation and finally, DSM considering battery degradation and HVAC. We are running this numerical simulation using a large office building, with 10 kWh of battery and 10 kW inverter. 

\begin{figure}
\includegraphics[width=0.5\textwidth]{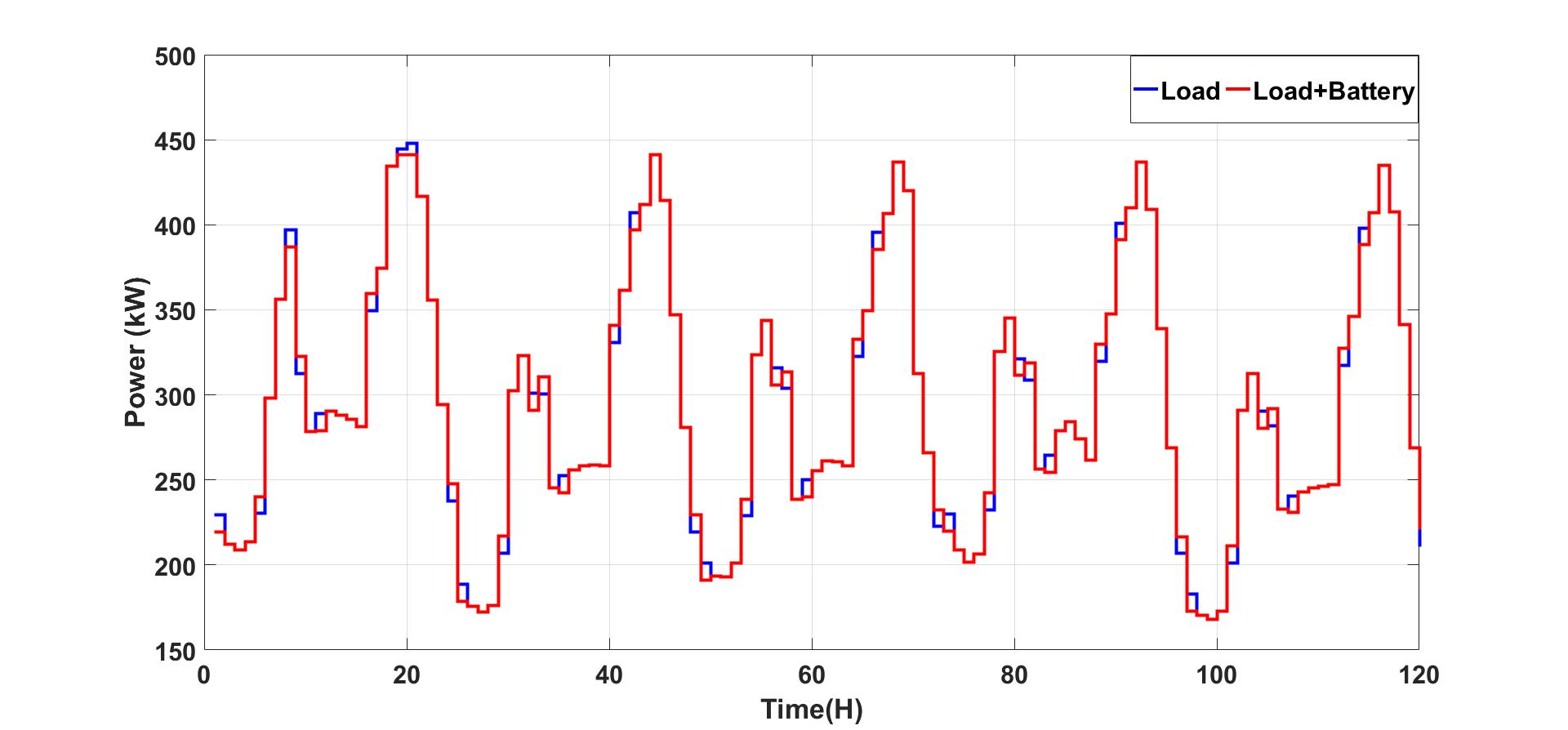}
\caption{DSM neglecting battery degradation}
\label{fig:base}
\end{figure}

\textbf{Case 1: DSM Neglecting Battery Degradation}
First of all, we show the battery operation without considering the battery degradation in Fig.~\ref{fig:base}. It shows the five days that has the largest peak loads throughout a billing cycle. It is observed that the battery operates in most each day and operates for multiple times. The annual saving and battery degradation are tabulated in Table~\ref{tab:base}. As shown in the Table ~\ref{tab:base}, the degradation cost is much higher than the annual savings. In other words, the battery life degrades very fast, the annual saving cannot cover the cost of battery degradation. This a very important finding. A lot of the DSM researches for battery integration do not account for battery degradation, which is not realistic.

\begin{table}
\centering
\begin{threeparttable}
    \caption{DSM neglecting battery degradation}\label{tab:base}
    \centering
     \begin{tabular}[ht\textwidth]{c c} 
     \hline
    annual saving(\$)\tnote{*}&annual degradation(\$)\tnote{*}\\
    \hline
    2,738/2,473&12,276/12,275\\
    \hline
    \end{tabular}
    \begin{tablenotes}
    \small
    \item[*] values in upper bound/lower bound.     
    \end{tablenotes}
    
    \end{threeparttable}
\end{table}

\begin{figure}
\includegraphics[width=0.5\textwidth]{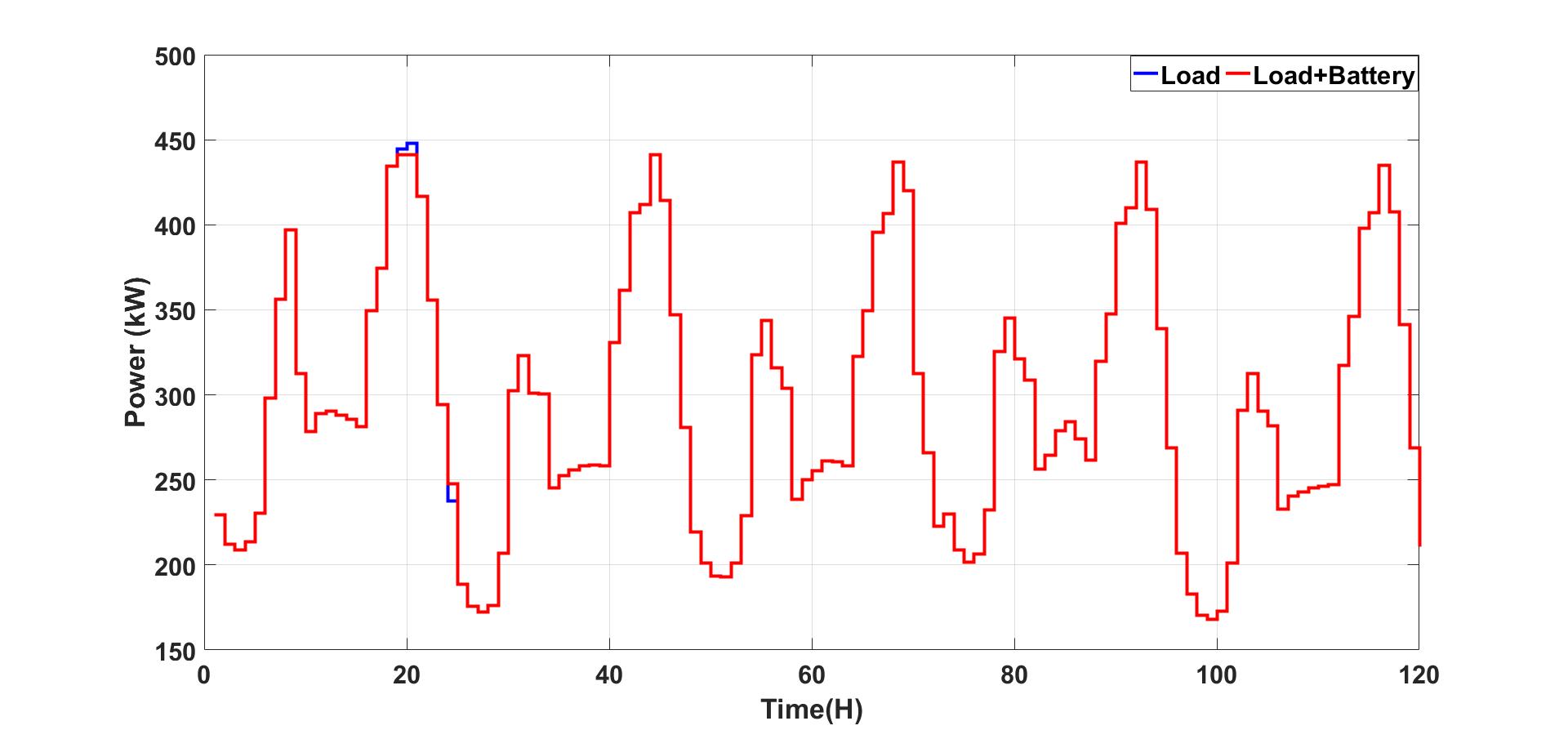}
\caption{DSM considering battery degradation}
\label{fig:withbat}
\end{figure}

\textbf{Case 2: DSM Considering Battery Degradation}
In the second case, we consider the operation of battery with the awareness of battery degradation. Similarly, we plot out the peak five days. As shown in Fig.~\ref{fig:withbat}, the battery only operates on the peak day, where the cut down of peak power can compensate for the battery degradation. Table ~\ref{tab:withbat} tabulates the annual savings and annual degradation cost. The saving from the battery integration could payback the battery in a relatively short time.

\begin{table}
	\centering
	\begin{threeparttable}
    \caption{DSM considering battery degradation}\label{tab:withbat}
    \centering
     \begin{tabular}[ht\textwidth]{c c} 
     \hline
    annual saving(\$)\tnote{*} &annual degradation(\$)\tnote{*}\\
    \hline
    2,461/2,086&317/234\\
    \hline
    \end{tabular}
    \begin{tablenotes}
    \small
    \item[*] values in upper bound/lower bound.     
    \end{tablenotes}
    
    \end{threeparttable}
\end{table}

\begin{figure}
\includegraphics[width=0.5\textwidth]{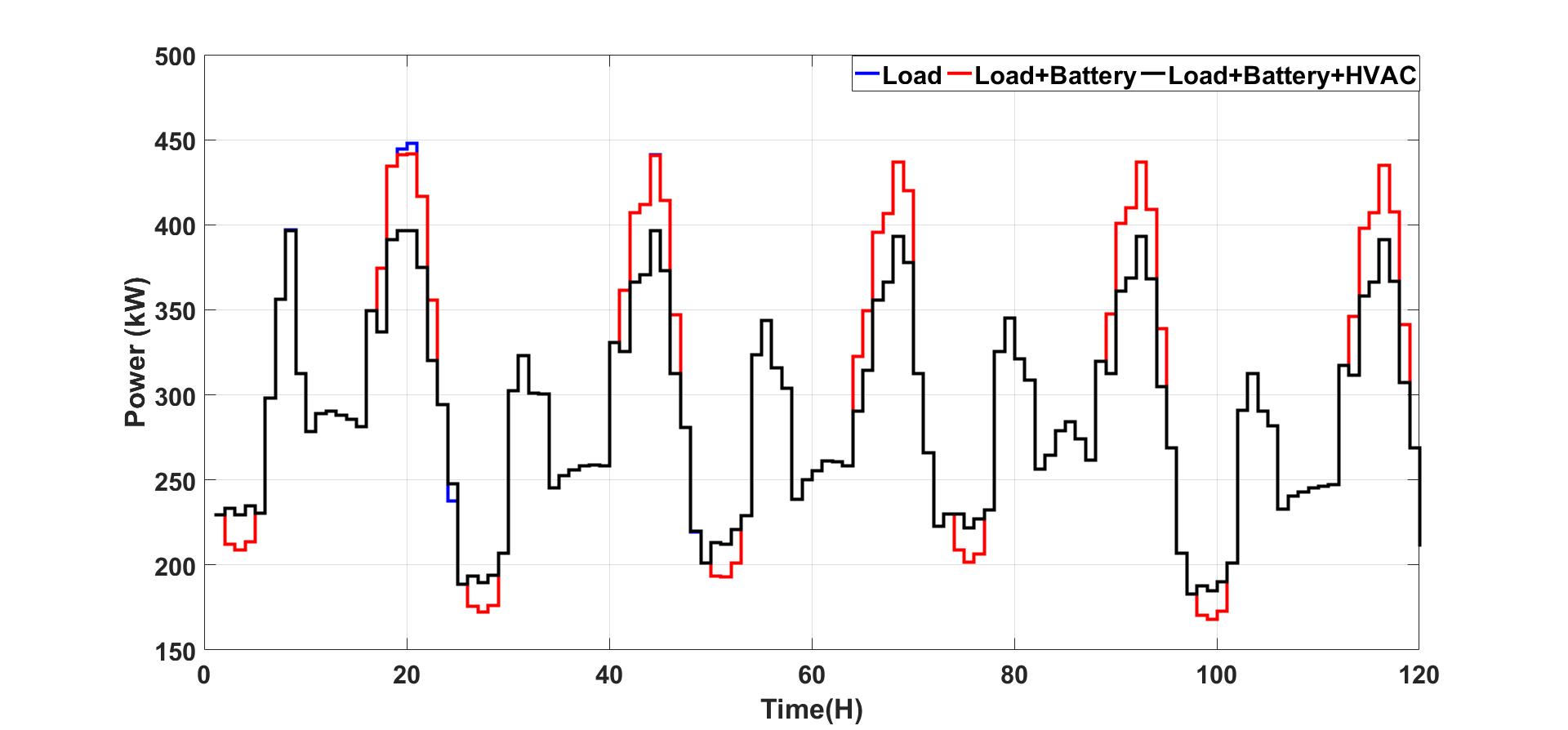}
\caption{DSM considering battery degradation and HVAC}
\label{fig:HVAC}
\end{figure}

\textbf{Case 3: DSM Considering Battery Degradation and HVAC}
In the last case, we further extend our study by integrating the HVAC control. The idea is to in corporate HVAC control so that it cools down the building in a pre-cooling period, and avoid the peak in peak hours. Similarly, we plot the first five peak days in the Fig.~\ref{fig:HVAC}. The annual savings and annual battery degradation are documented in Table ~\ref{tab:HVAC}. It is shown in the table that the saving is significant, while the battery degradation is at largely at the same value as Case 2. The saving is largely contributed to control of the HVAC. Numerical results show that by integrating HVAC control and battery to the commercial building, it significantly saves the electricity bill for building owners. 

\begin{table}
	\centering
	\begin{threeparttable}
    \caption{DSM considering battery degradation and HVAC}\label{tab:HVAC}
    \centering
     \begin{tabular}[ht\textwidth]{c c} 
     \hline
    annual saving(\$) \tnote{*}&annual degradation(\$) \tnote{*}\\
    \hline
    9,995/8,223&335/237\\
    \hline
    \end{tabular}
    \begin{tablenotes}
    \small
    \item[*] values in upper bound/lower bound.     
    \end{tablenotes}
    
    \end{threeparttable}

\end{table}

\subsection{Building Variations}
\begin{figure}
\centering
\includegraphics[width=0.45\textwidth]{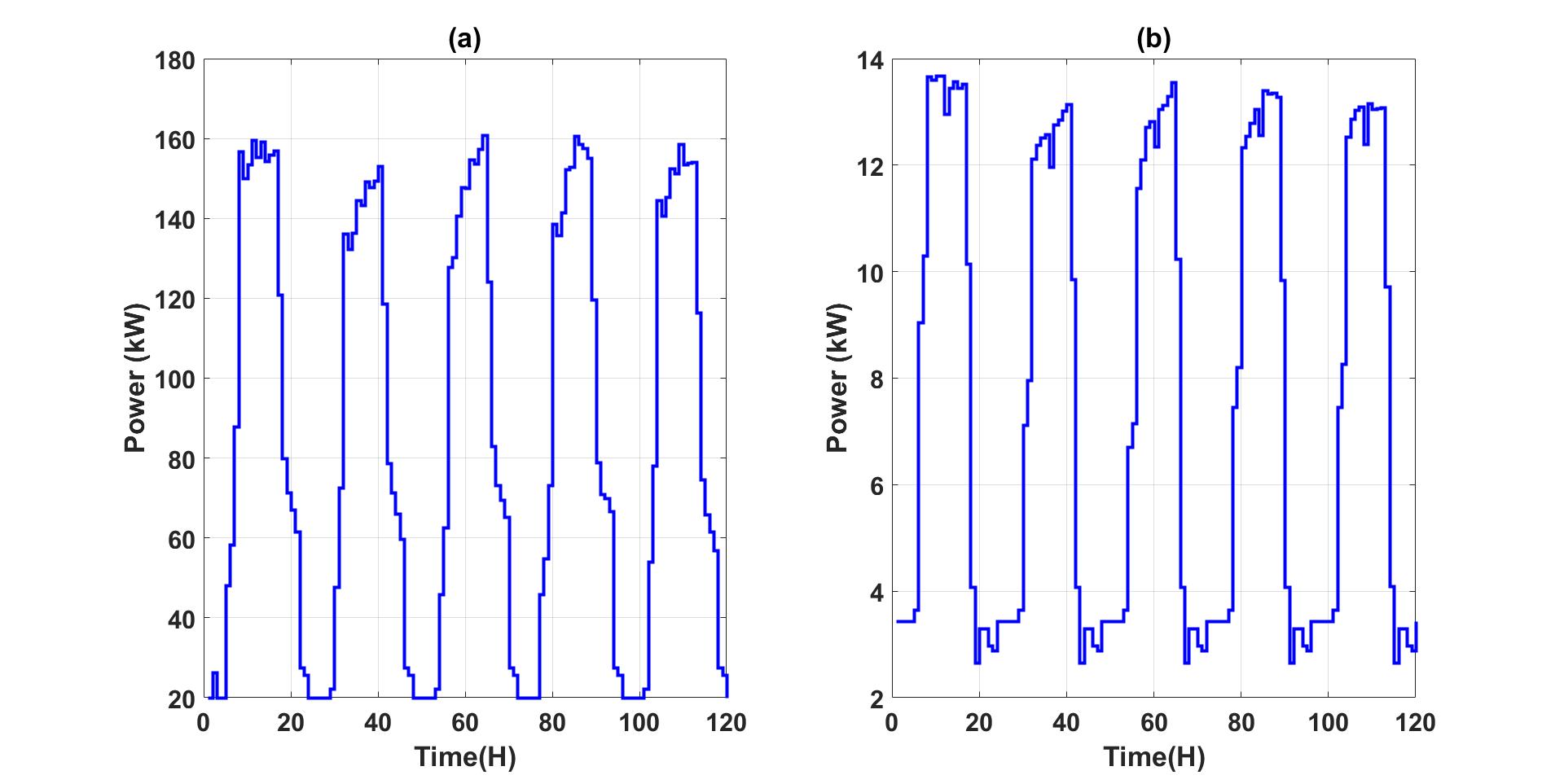}
\caption{Load pattern of different buildings (a) medium office (b) small office}
\label{fig:loadvar}
\end{figure}

To further validate our proposed methodology, we perform numerical experiments on different building types. The small and medium building load samples are plotted in Fig.~\ref{fig:loadvar}. We run the battery integration algorithm proposed in this paper and the results are tabulated in Table ~\ref{tab:loadvar}. The table shows the battery capacity, inverter size, annual saving, battery degradation, payback time and salvage value for the battery at the payback time. It is shown in the table that, the large office has the most potential for battery integration. The 2 year payback time is an optimal estimation.

\begin{table*}[t]
\centering
\begin{threeparttable}
    \caption{Comparisons of battery integration with different building}\label{tab:loadvar}
    
     \begin{tabular}{c|c c c c c} 
    \hline
    building type&battery(kWh)/inverter size(kW)&saving(\$)\tnote{*}&degradation(\$)\tnote{*} &payback time(year)\tnote{*}&battery salvage value(\%)\tnote{*}\\
     
    \hline
    large office&10。0/10.0&2,461/2,086&317/243&2.0/2.4&0.82/0.81\\
    medium office&5。0/5.0&1,058/906&118/78&2.4/2.8&0.81/0.86\\
    small office&2.0/2.0&185/169&24/24&5.4/5.8&0.78/0.76\\
    \hline
    \end{tabular}
    \begin{tablenotes}
    \small
    \item[*] values in upper bound/lower bound.      
    \end{tablenotes}
\end{threeparttable}
\end{table*}

\subsection{Climate Variations}
Climate as an important To further validate our proposed methodology, we perform numerical experiments and the results are tabulated in Table~\ref{tab:weather}. We have selected 9 large cities in 9 climate zones in U.S defined by ASHREE (excluding NYC in Table~\ref{tab:weather}). The results show that although climate changes from coast to coast, the developed design phase assessment tool works well for all regions across U.S.

\begin{table*}[t]
\centering
\begin{threeparttable}
    \caption{Comparisons of battery integration under different climate}\label{tab:weather}
    
     \begin{tabular}{c|c  c c c} 
    \hline
    weather zone&saving(\$)\tnote{*}&degradation(\$)\tnote{*} &payback time(year)\tnote{*}&battery salvage value(\%)\tnote{*}\\
     
    \hline
    Miami, FL&2,640/2,327&278/276&1.9/2.1&0.82/0.80\\
    Houston, TX&2,644/2,297&345/311&1.9/2.2&0.78/0.77\\
    Memphis, TN&2,642/2,304&338/306&1.9/2.2&0.79/0.78\\
    Baltimore, MD&2,478/2,116&331/240&2.0/2.4&0.77/0.81\\
    Chicago, IL&2,661/2,340&354/349&1.9/2.1&0.78/0.75\\
    Burlington, VT&2,569/2,232&335/297&1.9/2.2&0.78/0.78\\
    Duluth, MN&2,687/2,324&443/392&1.9/2.2&0.73/0.72\\
    Fairbanks,AK&2,685/2,305&485/414&1.9/2.2&0.70/0.70\\
    
    \hline
    \end{tabular}
    \begin{tablenotes}
    \small
    \item[*] values in upper bound/lower bound.
  
    \end{tablenotes}
\end{threeparttable}
\end{table*}

\subsection{Runtime Control}
\begin{figure}[!th]
\centering
\includegraphics[width=0.4\textwidth]{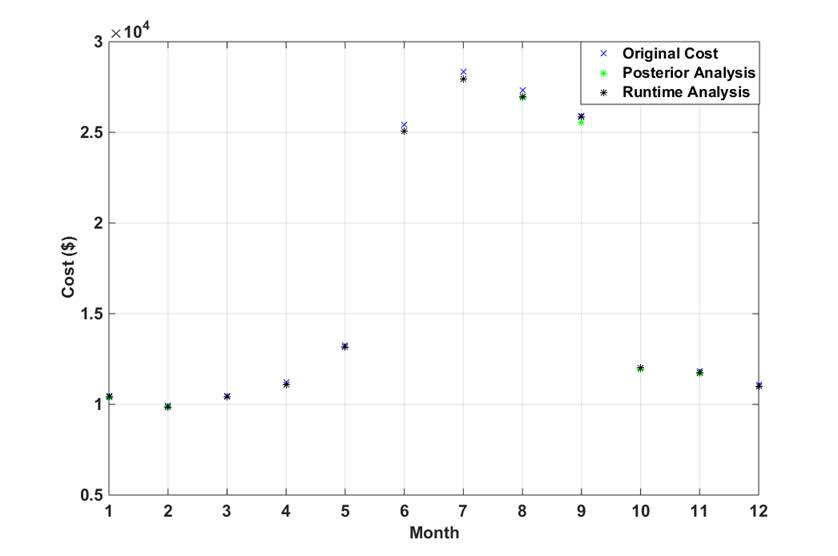}
\caption{Runtime saving versus design phase estimations}
\label{fig:runtime}
\end{figure}

We use the runtime control algorithm developed in the previous section for runtime control. We use the weather information of NYC for 2014 as the data source for sample average approximation, and apply 2015 weather patterns for verification. Fig.~\ref{fig:runtime} shows the comparisons of operating cost throughout the year between design phase estimations and runtime. It is demonstrated that most of the cost are in summer time, where electricity is primarily used for cooling. Table~\ref{tab:runtime} further compares the savings between the design phase and runtime. It is shown that even without perfect knowledge of the load profile, the runtime controller is able to achieve roughly 77\% of the total savings achieve by design phase optimum case. Looking back to the 2.5 years optimal payback time in Table \ref{tab:weather}, this implies that, in runtime, the payback time of the battery is around 3 years. Note the runtime control results could be further improved by incorporating more historical data.

\begin{table}
\centering
\begin{threeparttable}

    \caption{Runtime saving versus design phase estimations}\label{tab:runtime}
    \centering
     \begin{tabular}[ht\textwidth]{c c} 
     \hline
   design phase saving(\$) \tnote{*} & runtime saving (\$) \tnote{*} \\
    \hline
    2,251/1,929&  1,737/1,488\\
    \hline
    \end{tabular}
    \begin{tablenotes}
    \small
    \item[*] values in upper bound/lower bound.
    \end{tablenotes}
\end{threeparttable}
\end{table}

\section{Conclusions and Future Work}
In this paper, we have presented two algorithms for minimizing the payback time of battery integration to buildings. The two algorithms covers the both the design phase assessment and runtime control. We have utilized real-life tariff, commercially available battery and benchmark building loads for verification of the proposed algorithm. Algorithm 1 provides the best payback time for battery installation. Algorithm 2 is used on controllers and shows the real-life performance/payback time. Our simulation-based studies show that the battery system payback time can be as short as 3 years for representative commercial buildings in NYC. This methodology is applicable to commercial buildings at wide range of weather zones and/or under different tariff models. We have further demonstrated that the runtime control results match closely to the design phase  assessment, meaning the battery owner should expect a similar payback time to the time they plan battery integration.

There are several important findings in this paper: First, it make little sense to neglect battery degradation when designing DSM algorithms. The battery will wear out way before the payback time. Second, in most of the utilities, CPP is also time based. It is then hard to cast the tariff model as a convex problem. However, we have showed that it could be relaxed to bounds that are still narrow enough while granting us much faster speed. In the end, a key takeaway is battery integration to commercial buildings could have a short payback time. Deep discharge damages battery. However, if we only perform deep discharge a couple of times throughout the billing cycle wisely, we could incorporate a small initial investment (small battery installation) together with short payback time and high salvage value. 

In future, we plan to use sensor fusion for stochastic programming based runtime algorithm. As a constrain of this paper, we have not considered an extensive model of battery degradation due to its complexity. Thus, another direction is to incorporate a more detailed modeling of battery degradation.

\section*{Acknowledgment}
The authors appreciate the discussions with Mr. Tony Abate at NYSERDA, Prof. Christoph Meinrenken at Columbia University and Prof. Michael Bobker at City College of New York.
\section*{References}
\bibliographystyle{unsrtnat}
\bibliography{refNew}
\end{document}